\documentclass[aps,pre,preprint,groupedaddress,longbibliography]{revtex4-1}

\usepackage{amsmath}
\usepackage{amssymb}
\usepackage{bm}
\usepackage{booktabs}
\usepackage{capt-of}
\usepackage{epstopdf}
\usepackage{float}
\usepackage{graphicx}
\usepackage[table]{xcolor}
\usepackage[running]{lineno}
\usepackage{setspace}
\usepackage{textcomp}

\newcommand{\eg}{\citep[e.g.,][]}
\newcommand{\cf}{\citep[cf.][]}
\newcommand{\textunderscript}[1]{$_{\text{#1}}$}

\begin{document}

\title{A Heuristic Framework for Next-Generation Models of Geostrophic Convective Turbulence}

\author{Jonathan S.~Cheng}
\affiliation{Turbulence and Vortex Dynamics group, Department of Applied Physics and J.M. Burgers Center for Fluid Dynamics, Eindhoven University of Technology, Eindhoven, Netherlands}
\author{Jonathan M.~Aurnou}
\affiliation{Department of Earth, Planetary and Space Sciences, University of California, Los Angeles, Los Angeles, CA, USA}
\author{Keith~Julien}
\affiliation{Department of Applied Mathematics, University of Colorado at Boulder, Boulder, CO, USA}
\author{Rudie P.J.~Kunnen}
\affiliation{Turbulence and Vortex Dynamics group, Department of Applied Physics and J.M. Burgers Center for Fluid Dynamics, Eindhoven University of Technology, Eindhoven, Netherlands}

\begin{abstract}
Many geophysical and astrophysical phenomena are driven by turbulent fluid dynamics, containing behaviors separated by tens of orders of magnitude in scale. While direct simulations have made important strides toward understanding geophysical systems, such models still inhabit modest ranges of the governing parameters that cannot be extrapolated to planetary settings with confidence. The canonical problem of rotating Rayleigh-B\'enard convection provides an alternate approach -- isolating the fundamental physics in a reduced setting where more extreme parameter ranges can be accessed. Theoretical studies and asymptotically-reduced simulations in rotating convection have unveiled a variety of flow behaviors likely relevant to natural systems, but still inaccessible to direct simulation. In lieu of this, several new large-scale rotating convection devices have been designed to characterize such behaviors in the laboratory at previously inaccessible parameter values. With a potential influx of new data, it is essential to predict how upcoming results will fit into the network of existing results. Surprisingly, a coherent framework of predictions for extreme rotating convection has not yet been elucidated in the literature. In this study, we combine asymptotic predictions, results from laboratory and numerical data, and experimental constraints to build a heuristic framework for cross-comparison between a broad range of rotating convection studies. Diverse predictions exist for flow behavior and heat transfer in extreme rotating convection, originating from theory, asymptotically-reduced studies, direct numerical simulations, and laboratory experiments. We categorize these predictions in the context of asymptotic flow regimes, and discuss how to optimize laboratory experiments toward accessing asymptotic behaviors. We then consider the physical constraints that determine where the intersection between flow behavior predictions and experimental accessibility takes place. Applying this framework to several existing and upcoming experiments demonstrates that laboratory studies may soon be able to characterize geophysically-relevant flow regimes in rotating convection. These new laboratory data may transform our understanding of geophysical and astrophysical turbulence, and the conceptual framework developed herein should provide the theoretical infrastructure needed for meaningful discussion of these results.
\end{abstract}

\maketitle

\section{Introduction}

Turbulent flows underlie many geophysical and astrophysical phenomena in the universe, from the dynamics of the oceans and atmosphere on Earth to the fluid dynamos generating magnetic fields in planets and stars \eg{Marshall99, Miesch00, Heimpel05, RobertsKing13}. These flows are inherently difficult to investigate because their settings are too remote to allow for direct measurements. Thus, the main method for examining many such flows is to develop forward models \eg{Busse00, Bahcall01, Heimpel05, Monchaux07, Spence09, Jones11, Soderlund13}. Forward models aim to capture the underlying dynamics of geophysical systems in a simplified setting. Two common methods for modeling planetary physics are to directly simulate the governing flow equations using numerical models, or to investigate fluid behaviors in a laboratory setting. While direct numerical simulations more faithfully model the overall geometry and orientation of force vectors in a geophysical system \citep{Gastine15, Gastine16}, laboratory experiments can approach more extreme, geophysically-relevant conditions \citep{Niemela00, He14, Jones14, Aurnou15, Cheng15, Nataf15}.

Of the many forces involved in geophysical and astrophysical fluid processes, buoyant instabilities and rotational effects are often dominant. A reduced problem deeply relevant to these processes, then, is plane layer thermal convection under the influence of rotation. This canonical approach takes advantage of an extensive literature of Rayleigh-B\'{e}nard convection and rotating convection studies, including theory, direct numerical simulations (DNS), and laboratory experiments \eg{Malkus54, Rossby69, Julien96, Grossmann00, Ahlers09}. Some recent numerical models take a unique approach to rotating convection by solving modified governing equations in the limit of asymptotically rapid rotation \citep{Sprague06, JulienPRL12, Julien16, Plumley16}. Predictions from theory and from these `asymptotically-reduced' models have established that many of the behavioral regimes which are likely relevant to planetary-scale flows cannot yet be accessed by direct models of geophysical systems \citep{JulienGAFD12, Aurnou15}. The simpler geometry of the rotating convection problem is better suited for reaching parameter ranges where these regimes are expected to manifest \eg{Favier14, Guervilly14, Cheng15, Kunnen16}.

The Rayleigh number $Ra = \gamma g \Delta T H^{3} / \left( \nu \kappa \right)$ describes the strength of the buoyancy forcing in convection as the squared ratio between the viscous diffusion and thermal diffusion time scales, $\tau_\nu$ and $\tau_\kappa$, and the buoyancy forcing (free-fall) time scale $\tau_\mathit{ff}= H/U_\mathit{ff}$ squared. The convective free-fall velocity is defined here as $U_\mathit{ff} = \left(\gamma g \Delta T H\right)^{1/2}$, so $\tau_\mathit{ff} = H^{1/2} \left( \gamma g \Delta T \right)^{-1/2}$. The coefficient of thermal expansion is $\gamma$, $g$ is the gravitational acceleration, $\Delta T$ is the adverse superadiabatic temperature gradient, $H$ is the height of the fluid layer, $\nu$ is the kinematic viscosity and $\kappa$ is the thermal diffusivity. The Prandtl number $Pr = \nu / \kappa$ gives the ratio between the thermal and viscous diffusion timescales. The Ekman number, $E = \nu / \left( 2 \Omega H^{2} \right)$ parametrizes the influence of rotation as the ratio between the rotational time scale $\tau_\Omega = 1 / \left( 2 \Omega \right)$ and the viscous time scale, where $\Omega$ is the angular rotation rate of the body. The Rossby number $Ro = U / \left( 2 \Omega H \right)$ also describes the influence of rotation by comparing the rotational time scale $\tau_\Omega$ to the inertial time scale $\tau_i=H/U$, where $U$ is the characteristic flow velocity. In convectively-driven flows, the buoyant free-fall time scale $\tau_\mathit{ff}$ serves as a lower bound on the inertial time scale $\tau_i$ when all heating power goes toward fluid motions \eg{Gilman77, Julien96, Stevens09}. For the limit of $\tau_i = \tau_\mathit{ff}$, a `convective Rossby number' can be defined:
\begin{equation}
\label{eq:RoC}
Ro_C = \frac{\tau_{\Omega}}{\tau_\mathit{ff}} = \left( \frac{\gamma g \Delta T}{(2 \Omega)^2 H} \right)^{1/2} = \left( \frac{Ra E^2}{Pr} \right)^{1/2} \, .
\end{equation}

In geophysical settings these parameters take on extreme values -- in the Earth's outer core, for example, estimates give $Ra \sim 10^{20}-10^{30}$, $E \sim 10^{-15}$ and $Ro \sim 10^{-6}$ \citep{Gubbins01, Aurnou03, Schubert11}. A massive separation exists between the viscous and inertial time scales, as well as between the inertial and rotational time scales ($\tau_\nu \gg \tau_i \gg \tau_\Omega$). The majority of direct simulations of the outer core, in contrast, are confined to ranges of $Ra \lesssim 10^7$, $E \gtrsim 10^{-6}$, and $Ro \gtrsim 10^{-2}$ due to numerical resolution constraints \eg{KingBuffett13, Sreenivasan14}. Many of the behaviors expected from theory and asymptotic models may not yet manifest in these models.

\begin{figure}
\centering
\includegraphics[width=1.0\linewidth]{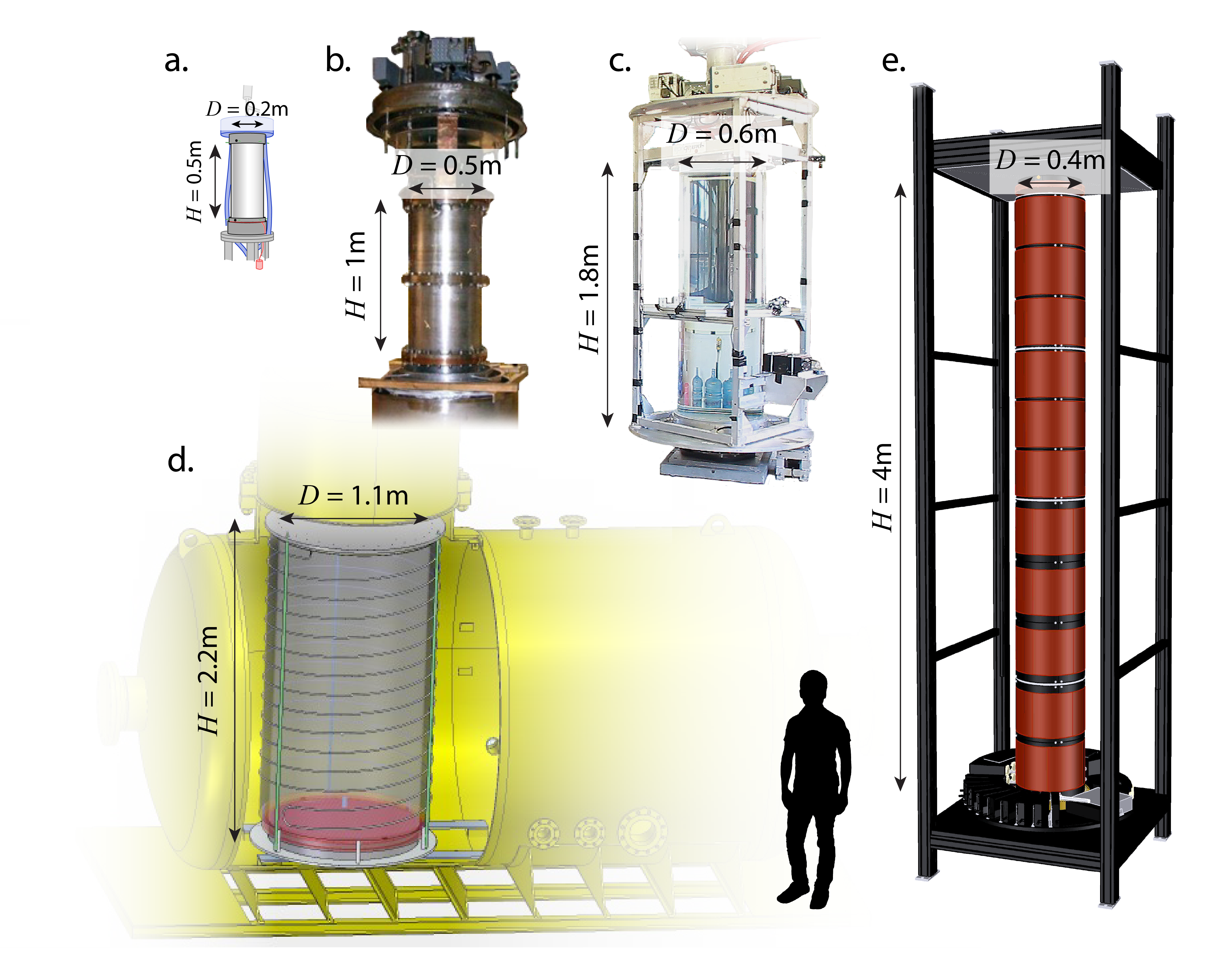}
\caption{\label{F:imgs} \footnotesize Images of several extreme rotating convection setups. a) `RoMag' at UCLA (liquid gallium, $Pr \approx 0.025$) \cite{KingAurnou13}. b) Trieste experiment at ICTP (cryogenic liquid He, $Pr \approx 0.7$) \cite{Niemela00, Ecke14}. c) `NoMag' at UCLA (water, $Pr \approx 4 - 7$). d) `U-Boot' at the Max Planck Institute for Dynamics and Self-Organization (SF\textunderscript{6}, N\textunderscript{2}, He gas, $Pr \approx 0.8$) \cite{AhlersNJP09, Funf09}. e) `TROCONVEX' at Eindhoven University of Technology (water, $Pr \approx 2 - 7$).
}
\end{figure}

While numerical models need to resolve the scale separation between different behaviors to simulate geophysically meaningful flows, laboratory experiments inherently `resolve' all of the physics, even for behaviors that are too small to detect \citep{RobertsKing13}. The laboratory approach is thus uniquely well-suited for investigating geophysical-style rotating convection at extreme values of the governing parameters. Several new large-scale rotating convection devices have been built for this purpose (Figure \ref{F:imgs}). With a potential influx of new experimental data, it is crucial to build theoretical infrastructure that fosters cross-comparison between different experiments.

In this study, we construct a predictive framework by combining asymptotic results for geostrophic convection regimes, regime transitions and heat transfer scalings detected in existing laboratory and numerical studies, and physical constraints governing the accessible parameter ranges for laboratory experiments. Design considerations for optimizing experiments toward exploring extreme parameter ranges are also discussed. We show that upcoming experiments may be able to reach conditions where asymptotically-predicted flow behaviors manifest, allowing us to determine the relevance of such behaviors to natural phenomena.

In Section \ref{regime}, we review the behavioral regimes found in theoretical studies of rotating convection as well as the heat transfer scalings and flow transitions observed so far in laboratory experiments and DNS. In Section \ref{optim}, the design considerations for laboratory experiments to access and characterize these regimes are outlined. In Section \ref{constr}, we discuss the rotational and heat transfer constraints needed for ensuring that the physics remains within the bounds of classical Boussinesq rotating convection. To contextualize these experimental considerations with respect to flow regime predictions, we detail the achievable parameter ranges in a collection of extreme rotating convection devices (pictured in Figure \ref{F:imgs}). The significance of such laboratory experiments toward understanding geophysical systems is discussed in Section \ref{disc}.

\section{Flow regimes}
\label{regime}

Results from laboratory experiments, direct numerical simulations, and asymptotically-reduced studies indicate that a variety of rotating convection flow regimes occupy the range between rotationally-controlled and buoyancy-controlled convection \citep{JulienGAFD12}. To analyze the capabilities of a given experiment, we first outline these regimes and the parameter ranges over which they are expected to arise.

One method for categorizing flow behavior is to track the strength of the nondimensional heat transfer: different behaviors likely lead to different modes of heat transport, and thus to differences in the scaling properties \eg{Malkus54, Kraichnan62, Castaing89, Julien96, Grossmann00, Ahlers09}. The Nusselt number $Nu = q H / \left( \kappa \rho C_p \Delta T \right)$ represents the ratio between the total heat transfer and conductive heat transfer, where $C_p$ is the specific heat capacity and $q$ is the heat flux per unit area \eg{ChengAurnou16}. The governing parameters tend to be approximately related by power law scalings \eg{Spiegel71}; \cf{Grossmann00}. Here, we assume that the Nusselt number scales as:
\begin{equation}
\label{eq:NuRa}
Nu \sim Ra^{\alpha} E^{\beta} Pr^{\delta}\, ,
\end{equation}
where $\alpha$, $\beta$ and $\delta$ are constant exponents in a given scaling regime. Fundamental behavioral transitions are associated with changes in the mode of heat transfer and therefore with changes in these scaling exponents \eg{JulienGAFD12, Ecke14, Stellmach14, Cheng15, Kunnen16}.

\begin{figure}
\centering
\includegraphics[width=0.6\linewidth]{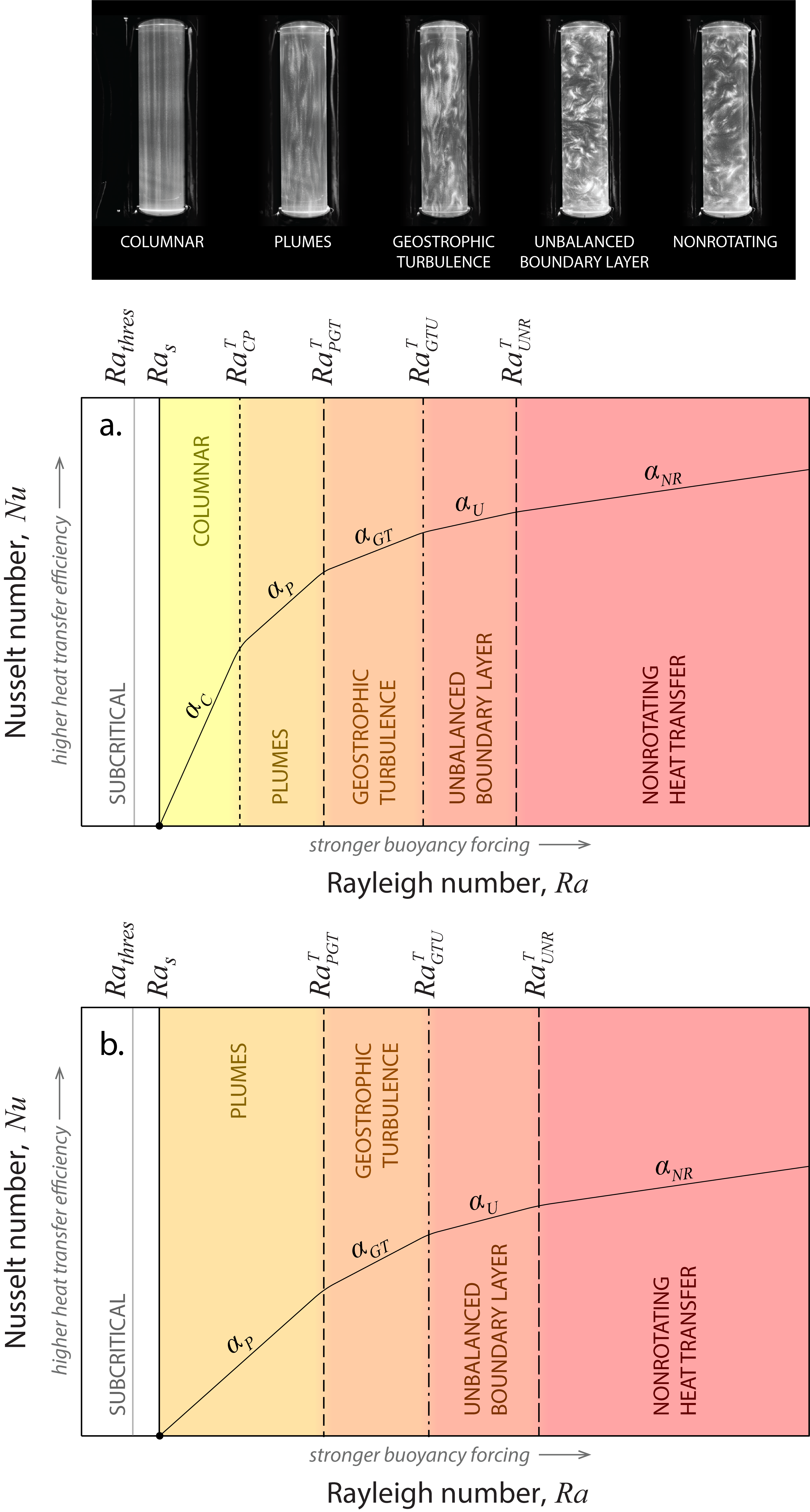}
\caption{\label{F:NuRa} \footnotesize Schematic showing the distribution of rotating convection regimes in terms of Nusselt number ($Nu$) versus Rayleigh number ($Ra$) for a fixed Ekman number ($E$) and a) $Pr > 3$ and b) $Pr \lesssim 3$. Laboratory flow visualizations of each regime at $Pr \approx 7$, originally published in \citet{Cheng15}, are shown in the upper panel. In a) and b), the vertical lines indicate transition Rayleigh values: $Ra_{s}$ denotes convective onset, $Ra_{_\mathit{CP}}$ denotes the transition between columnar-style convection and plumes, $Ra_{_\mathit{PGT}}$ between plumes and geostrophic turbulence, $Ra_{_\mathit{GTU}}$ between geostrophic turbulence and unbalanced boundary layers, and $Ra_{_\mathit{UNR}}$ to nonrotating-style convection. Though the transitions are delimited by lines, each likely occurs gradually over a range of $Ra$ values. Their locations are not yet well-determined, and Table \ref{tab:RaT} and Figure \ref{F:RaE} list various existing predictions. For $Pr \lesssim 3$, steady columnar convection does not occur \eg{JulienGAFD12, Stellmach14}.}
\end{figure}

Figure \ref{F:NuRa} is a schematic demonstrating how $Nu$ scales with $Ra$, from the onset of convection at low $Ra$ to flows indistinguishable from nonrotating convection at high $Ra$ (and assuming fixed $E$ and $Pr$ values). The flow changes morphology a number of times between these endpoints, resulting in multiple behavioral regimes. The `columnar', `plumes' and `geostrophic turbulence' regimes are derived from the asymptotic results of \citet{JulienGAFD12} and \citet{Nieves14} while properties of the `nonrotating heat transfer' regime are established in classical experiments and theory \eg{Malkus54, Kraichnan62, Spiegel71, Castaing89}. Note that between onset and $Ra/Ra_{s} \sim 2$, flow exists in the `cellular' regime \citep{Veronis59}. This regime is not marked separately on Figure \ref{F:NuRa}, as the heat transfer scaling does not change appreciably between this regime and the next \citep{JulienGAFD12}. We theorize that a regime of `unbalanced boundary layers' occurs beyond geostrophic turbulence but prior to the flow becoming fully insensitive to rotation. The flow behaviors of each regime are described in Appendix \ref{asym-regime}.

The columnar regime only appears for Prandtl numbers greater than 3 \eg{JulienGAFD12}. Hence, $Pr=3$ is used as an approximate threshold between `large' and `small' Prandtl numbers, and Figure \ref{F:NuRa}a shows the predicted regimes for $Pr > 3$ flows while \ref{F:NuRa}b shows the predicted regimes for $Pr \lesssim 3$ flows. In both cases, the $Nu$--$Ra$ scaling exponent $\alpha$ decreases as the convective forcing increases in strength relative to rotational effects. Literature predictions for $\alpha$ are described in Appendix \ref{alpha-regime}. Rotating convection at $Pr <  0.68$ again exhibits distinct behaviors \eg{Chandra61,Horn17}. The main text will concern only $Pr \gtrsim 0.68$ fluids, while a discussion of $Pr < 0.68$ fluids, particularly $Pr \ll 1$ fluids such as liquid metals, is allocated to Appendix \ref{metals}.

\begin{figure}
\centering
\includegraphics[width=0.65\linewidth]{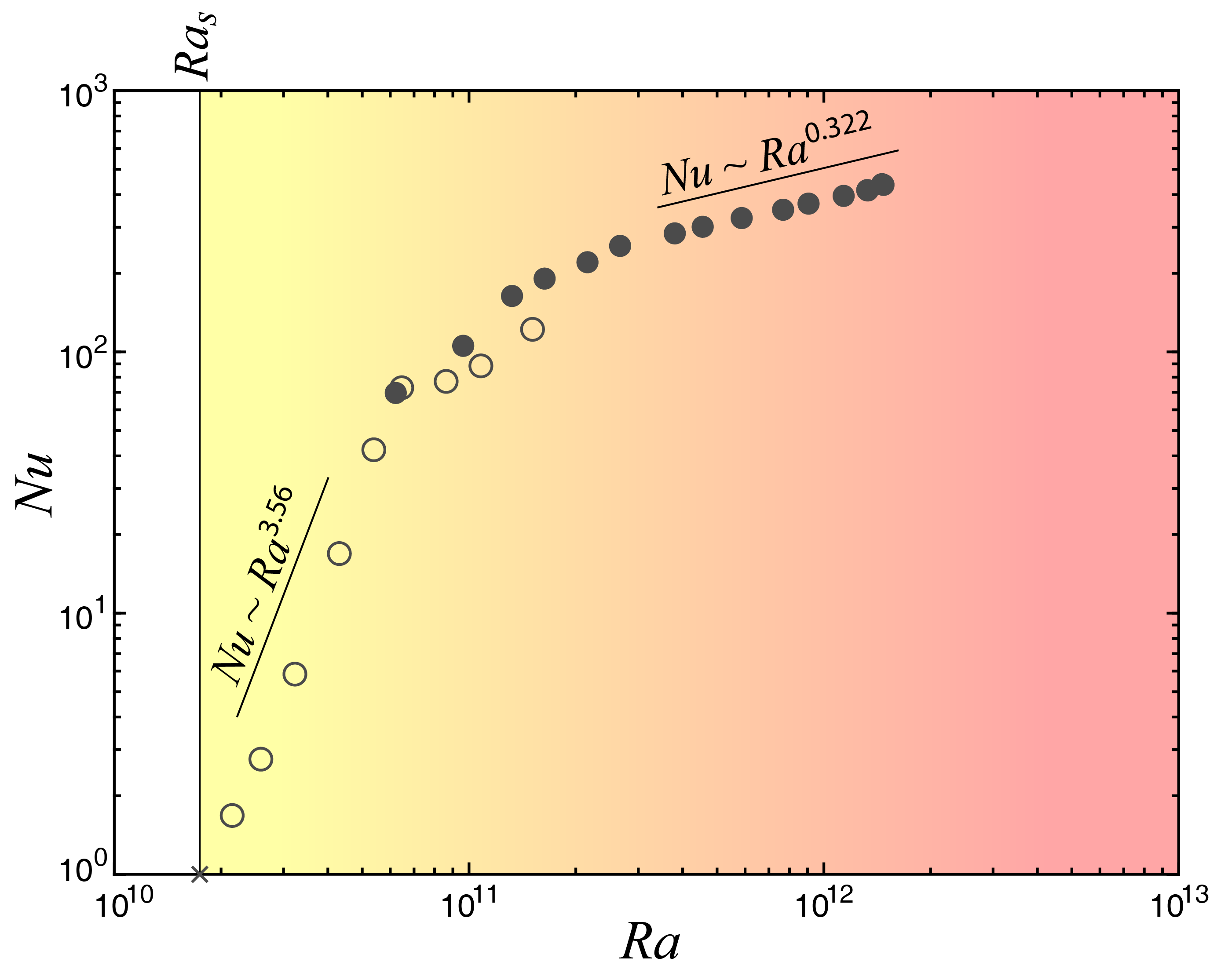}
\caption{\label{F:exNuRa} \footnotesize Example of $E \approx 10^{-7}$ rotating convection data, adapted from \citet{Cheng15}. Filled points correspond to laboratory experiments and open points correspond to DNS. The `$\times$' indicates the location of steady convective onset. A steep trend of $Nu \sim Ra^{3.56}$ occurs near onset while a shallow nonrotating convection trend of $Nu \sim Ra^{0.322}$ is approached at higher $Ra$ values. Between these two endpoints, there are no unambiguously distinguishable $Nu$--$Ra$ power law scalings.}
\end{figure}

Transition Rayleigh numbers $Ra_{_\mathit{CP}}$, $Ra_{_\mathit{PGT}}$, $Ra_{_\mathit{GTU}}$, and $Ra_{_\mathit{UNR}}$ separate the flow regimes in Figure \ref{F:NuRa}. While variations in $\alpha$ could function as a simple diagnostic for detecting regime transitions, most rotating convection studies find a relatively smooth transition in $\alpha$ between the endpoints of `rotationally-dominated' convection and `buoyancy-dominated' convection \eg{Rossby69, King12, Cheng15}, as shown in Figure \ref{F:exNuRa}. It is therefore important to give predictions for where regime transitions are expected in next-generation experiments. Fortunately, the rotating convection literature contains a wide variety of theoretical predictions and experimental results for regime transitions that may apply to the asymptotic schema. We compile transitions observed in the literature in Table \ref{tab:RaT} and, based on the physical arguments contained in the originating studies, we categorize them with respect to theoretically-predicted regimes. The properties of each transition and rationale for their categorization are described in more detail in Appendix \ref{RaT-regime}.

Steady, bulk convection onsets at a critical Rayleigh number $Ra_{s}$ in the form of overturning cells. For $Pr \gtrsim  0.68$ and $E \lesssim 10^{-3}$ \citep{Chandra61},
\begin{equation}
\label{eq:Ras}
Ra_{s} = 8.7 E^{-4/3} \, .
\end{equation}
Though $Ra_s$ is used to indicate onset in Figure \ref{F:NuRa}, the topic requires further discussion: in a finite container, instabilities often first occur as drifting waves attached to the sidewall of the container rather than as bulk motions. In the asymptotic case ($E \rightarrow 0$), these `wall modes' onset at \citep{Herrmann93, Zhang09}:
\begin{equation}
\label{eq:Raw}
Ra_{w} = 31.8 E^{-1} \, ,
\end{equation}
with vertical and azimuthal wavenumber $=1$ for cylindrical containers of aspect ratio $1/10 \leq \Gamma \leq 1$ (G. Vasil, private communications). Here, $\Gamma = D/H$ of the fluid layer where $D$ is the diameter. For the sake of simplicity, we will assume going forward that $Ra_s$ always corresponds to the onset of bulk convection \cf{Ecke15}. It is important to note that the fluid may become unstable to wall modes as early as $Ra_w \lesssim Ra_s/100$ for some of the experiments we discuss below. Thus, we cannot discount the possibility of wall mode-induced turbulence occurring in the bulk prior to stationary onset \citep{Horn17}. Though we will not address them here, open questions abound with regard to wall modes both at very low Ekman numbers and in low $\Gamma$ containers.

\begin{table*}
\caption{\label{tab:RaT}\footnotesize Table showing various predictions for the transitions between different flow regimes (shown schematically in Figures \ref{F:NuRa} and \ref{F:gen}). In the `Type' column, $Ra_{_\mathit{CP}}$ refers to the breakdown of well-organized convective columns into plumes, $Ra_{_\mathit{PGT}}$ refers to the breakdown of plumes into geostrophic turbulence, $Ra_{_\mathit{GTU}}$ refers to the local loss of rotational influence leading to unbalanced boundary layers and $Ra_{_\mathit{UNR}}$ refers to the global loss of rotational influence leading to nonrotating-style convection. The `$Pr$' column refers to the approximate Prandtl number for which the transition is observed or is predicted to apply. The `Reference' column gives the study from which each prediction originated. The `Figure abbreviation' column gives the label assigned to each transition in Figure \ref{F:RaE}. 
}
\begin{ruledtabular}
\begin{tabular}{@{}lllll}
Transition prediction & Type & $Pr$ & Reference & Figure abbr.\\
\hline
\rule{-2.8pt}{2.5ex} 
$Ra \sim 5.4 E^{-1.47}$ & $Ra_\mathit{CP}$ & $\approx 7$ & \citet{Cheng15} & $Ra_{\text{Ch15}}$\\[2pt]
\hline
\rule{-2.8pt}{2.5ex}
$Ra/Ra_{s} \sim 3$ & $Ra_\mathit{PGT}$ & $<3$ & \citet{JulienPRL12} & $Ra/Ra_{s} = 3$\\[2pt]
\hline
\rule{-2.8pt}{2.5ex}
$Ra \sim E^{-8/5} Pr^{3/5}$ & $Ra_\mathit{GTU}$ & any* & \citet{JulienPRL12} & $Ra_{\text{Ju12}}$\\
$Ra \sim 1.3 E^{-1.65}$ & $Ra_\mathit{GTU}$ & $\approx 6$ & \citet{Ecke14} & $Ra_{\text{EN14.1}}$\\
$Ra \sim 0.25 E^{-1.8}$ & $Ra_\mathit{GTU}$ & $\approx 0.7$ & \citet{Ecke14} & $Ra_{\text{EN14.2}}$\\[2pt]
\hline
\rule{-2.8pt}{2.5ex}
$Ro_C \sim 0.35$ & $Ra_\mathit{UNR}$ & $\approx 0.7$ & \citet{Ecke14} & $Ro_C=0.35$\\
$Ra \sim 100 E^{-12/7}$ & $Ra_\mathit{UNR}$ & $1$ & \citet{Gastine16} & $Ra_{\text{Ga16}}$\\
$Ro_C \sim 1$ & $Ra_\mathit{UNR}$ & any & \citet{Gilman77} & $Ro_C=1$\\
\end{tabular}
\end{ruledtabular}
\rule{0pt}{2.5ex}
\raggedright{ \footnotesize
*While \citep{JulienPRL12} did not reach the geostrophic turbulence regime for any $Pr>3$ cases, the asymptotic argument for this transition is $Pr$-independent.\\
}
\end{table*}

\section{Design factors}
\label{optim}

\begin{figure*}
\centering
\includegraphics[width=0.85\linewidth]{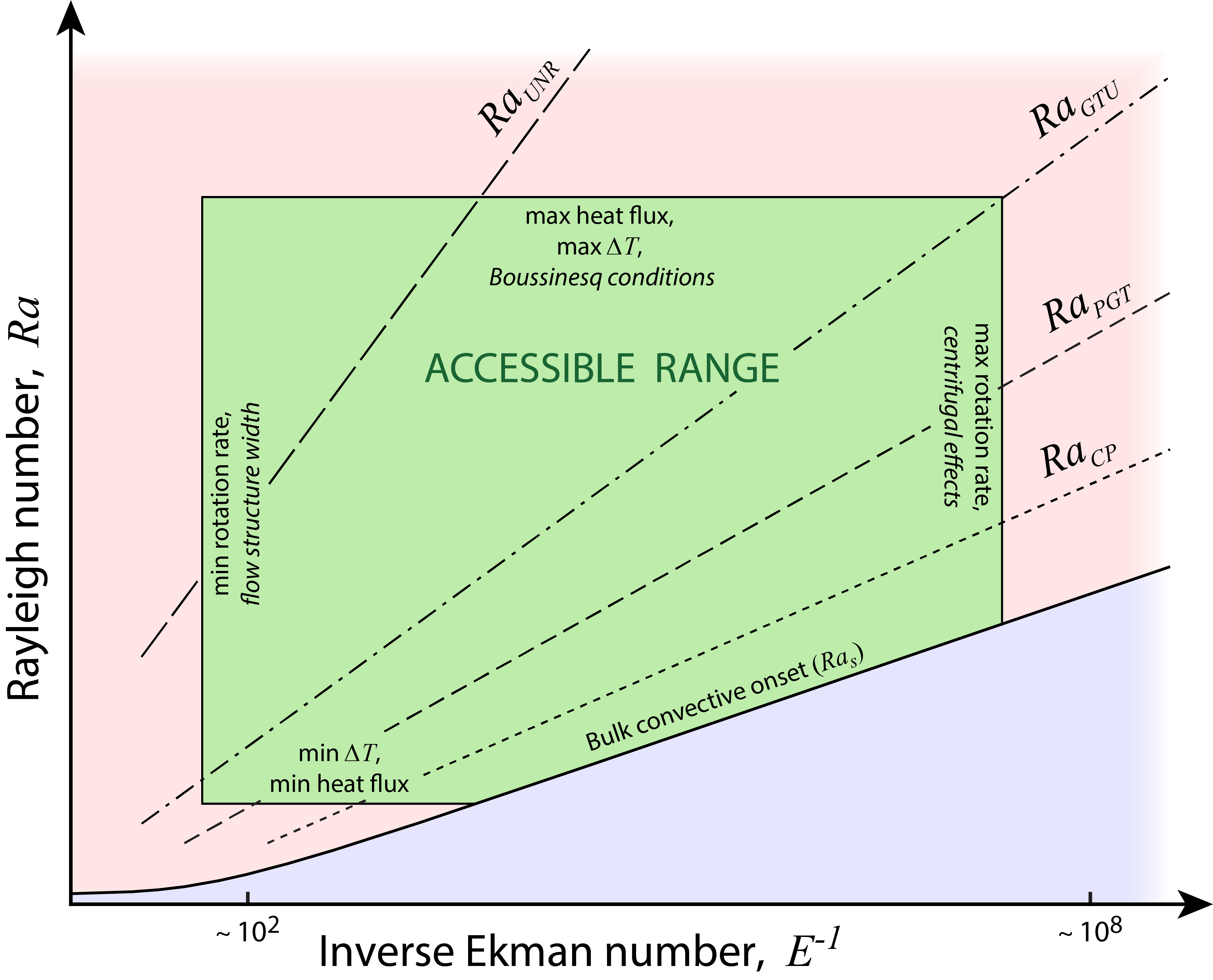}
\caption{\label{F:gen} \footnotesize Schematic of accessible Ekman number, $E$, and Rayleigh number, $Ra$ ranges for a given rotating convection experiment. Assuming the tank size and fluid properties are fixed, the absolute bounds on $E$ are determined by the minimum and maximum rotation rates, $\Omega$, and the absolute bounds on $Ra$ are determined by the minimum and maximum temperature difference, $\Delta T$. Minimizing sidewall effects and minimizing centrifugation effects require separate lower and upper bounds on $\Omega$. Maintaining Boussinesq conditions requires a separate upper bound on $\Delta T$. The onset of bulk convection, $Ra_{s}$ is indicated by a solid black line. The threshold for Ekman pumping effects on the heat transport is indicated by a solid grey line. Different flow regimes are separated by transition Rayleigh values $Ra_{_\mathit{CP}}$, $Ra_{_\mathit{PGT}}$, $Ra_{_\mathit{GTU}}$ and $Ra_{_\mathit{UNR}}$ (see Table \ref{tab:RaT}). These transitions and the regimes they separate are difficult to distinguish at moderate-to-high $E$ values but become distinct as $E$ decreases.}

\end{figure*}

Present-day laboratory and DNS studies are, at best, only partially able to capture the asymptotic behaviors we have catalogued above \cf{Favier14, Guervilly14, Stellmach14}. To further our understanding of extreme rotating convection, laboratory experiments must be optimized toward comparison with theory by covering broad ranges of $Ra/Ra_{s}$ at extremely low values of $E$ (e.g., $E < 10^{-7}$). Here, we discuss the physical considerations essential to designing these experiments.

Figure \ref{F:gen} is a schematic showing the accessible $E$ and $Ra$ ranges in a rotating convection setup with fixed height and width. Assuming the fluid properties are also fixed, the bounds on $E$ are determined solely by the minimum and maximum rotation rates of the system, $\Omega$, and the bounds on $Ra$ are determined solely by the minimum and maximum imposed temperature difference, $\Delta T$. The temperature difference is often imposed by applying a fixed heat flux $q$ to the bottom boundary \eg{AhlersNJP09, Ecke14, Cheng15}, meaning the control parameter is the flux Rayleigh number:
\begin{equation}
\label{eq:RaF}
Ra_F = Nu \cdot Ra = \frac{\gamma g H^4 q}{\nu \kappa k} \, ,
\end{equation}
where $k = \rho C_p \kappa$ is the thermal conductivity of the fluid. For a shallow $Nu$--$Ra$ scaling such as $\alpha_{_\mathit{NR}} \simeq 1/3$, $Ra \propto q^{3/4}$, while for a steep $Nu$--$Ra$ scaling such as $\alpha_{C} \simeq 3$, $Ra \propto q^{1/4}$. In both cases, varying $q$ is relatively inefficient for accessing broad ranges of $Ra$.

In contrast to the linear dependence of $\Delta T$ on $Ra$ and $\Omega$ on $E$, $Ra$ varies with $H^3$ and $E$ varies with $H^{-2}$: changing the height of the experiment is far more effective for reaching a broad range of $Ra$ and $E$ values \citep{Zhong91}. However, increasing the height simultaneously hinders the ability to access low values of $Ra/Ra_{s}$. The supercriticality is given by:
\begin{equation}
 \label{eq:supercrit}
\frac{Ra}{Ra_{s}}=\frac{Ra}{8.7 E^{-4/3}} = \frac{\gamma g \nu^{1/3}}{21.9 \kappa}\frac{\Delta T H^{1/3}}{\Omega^{4/3}} \, .
\end{equation}
Since $Ra/Ra_{s} \propto H^{1/3}$, a higher tank height corresponds to a higher minimum supercriticality, thus restricting the ability to access near-onset flow regimes. To overcome this limitation, the RoMag, NoMag, and TROCONVEX experiments (Figure \ref{F:imgs}a, c, d and e) use interchangeable tanks of various heights. Table \ref{tab:app} in Appendix \ref{exptinfo} contains more information about these experiments. In order to completely bridge the gap between onset and the minimum achievable $Ra/Ra_{s}$ in experiments, though, direct numerical simulations have proven to be ideal \citep{King12, Cheng15}.

The U-Boot and Trieste experiments (Figure \ref{F:imgs}b and d) can access high $Ra$ values by taking advantage of the large ratio between thermal expansivity and the thermal and viscous diffusivites ($\gamma/\nu\kappa$) in cryogenic helium and other compressed gases \citep{Niemela00, AhlersNJP09, Funf09, Niemela10}. For example, at a typical operating temperature and pressure for cryogenic helium of 4.7 K and 0.12 bar, $\gamma/\nu\kappa \approx 10^{11}$ s\textsuperscript2m\textsuperscript{-4}K\textsuperscript{-1}. At a typical operating temperature of 25 \textdegree C for water, $\gamma/\nu\kappa \approx 2 \times 10^{9}$ s\textsuperscript2m\textsuperscript{-4}K\textsuperscript{-1}, a factor of 50 below that of helium.

Furthermore, the ability to vary the pressure allows for a greater $Ra$-range in gas experiments. From the definition of $Ra$, we see that for an ideal gas:
\begin{equation}
\label{eq:Ragas}
Ra=\frac{\gamma g \Delta T H^3 \rho^2 C_P}{k \eta} \propto \rho^2 \propto P^2 M^2 \, ,
\end{equation}
where $\eta$ is the dynamic viscosity, $P$ is the pressure and $M$ is the molecular weight \citep{AhlersNJP09}. The quadratic relation between $Ra$ and pressure is especially useful since the pressure can be varied over several decades in these devices. The U-Boot device can also be filled with different gases of varying molecular weight in order to reach broader ranges of $Ra$ and lower values of $Ra/Ra_{s}$. Table \ref{tab:app2} lists some material properties for fluids used in each experiment.

\section{Experimental constraints}
\label{constr}

Additional limitations on the parameter coverage must be imposed to ensure that the fluid physics remains consistent with the fundamental rotating Rayleigh-B\'enard convection problem (indicated by italicized text in Figure \ref{F:gen}):

\begin{itemize}
\item Ensuring flow structures are not overly affected by tank geometry, which constrains the minimum rotation rate
\item Keeping the fluid in a Boussinesq state, which constrains the maximum imposed temperature gradient
\item Minimizing centrifugation effects, which constrains the maximum rotation rate
\end{itemize}
These arguments are compiled in Table \ref{tab:constr}, and we use the experimental devices shown in Figure \ref{F:imgs} as examples to demonstrate the effect on accessible $Ra$ and $E$ ranges.

\begin{table*}
\caption{\label{tab:constr}\footnotesize Table cataloging upper and lower bounds on the rotation rate ($\Omega$) and Ekman number ($E$), and the upper bounds on the temperature gradient ($\Delta T$) and Rayleigh number ($Ra$), for cylindrical rotating convection experiments. The constant prefactor $c=2.4$. The constraints on the flow structure width ratio ($m$) and the Froude number ($Fr$) are described in (\ref{eq:m}) and (\ref{eq:Fr}), respectively.
}
\begin{ruledtabular}
\begin{tabular}{@{}llll}
Condition & Dimensional constraint & Nondimensional constraint \\
\hline
\noalign{\vskip 3pt}
$m \geq 10$ & $\Omega_{\min} = \dfrac{500 c^3 \nu H}{D^3}$ & $E_{\max} = \left(\dfrac{D}{10 c H} \right)^3$ \\
\noalign{\vskip 3pt}
$Fr < 0.1$ & $\Omega_{\max} = \left(\dfrac{0.2 g}{D} \right)^{1/2}$ & $E_{\min} = \left(\dfrac{1.25 \nu^2 D}{H^4 g} \right)^{1/2}$ \\
\noalign{\vskip 5pt}
$\gamma \Delta T < 0.1$ & $\Delta T_{\max} = \dfrac{0.1}{\gamma}$ & $Ra_{\max} = \dfrac{0.1 g H^3}{\nu \kappa}$ \\
\noalign{\vskip 2pt}
\end{tabular}
\end{ruledtabular}
\end{table*}

In an experimental setup, the tank geometry can affect the physics appreciably \eg{Wu92}. Many low-$E$ numerical simulations use doubly-periodic horizontal boundary conditions with effectively no walls \cf{Julien18}. Limiting sidewall effects in experiments is therefore important for ensuring valid comparisons with DNS. To this end, we implement the criterion that a large number of flow structures must fit horizontally across the tank.

\begin{figure*}
\centering
\includegraphics[width=1.0\linewidth]{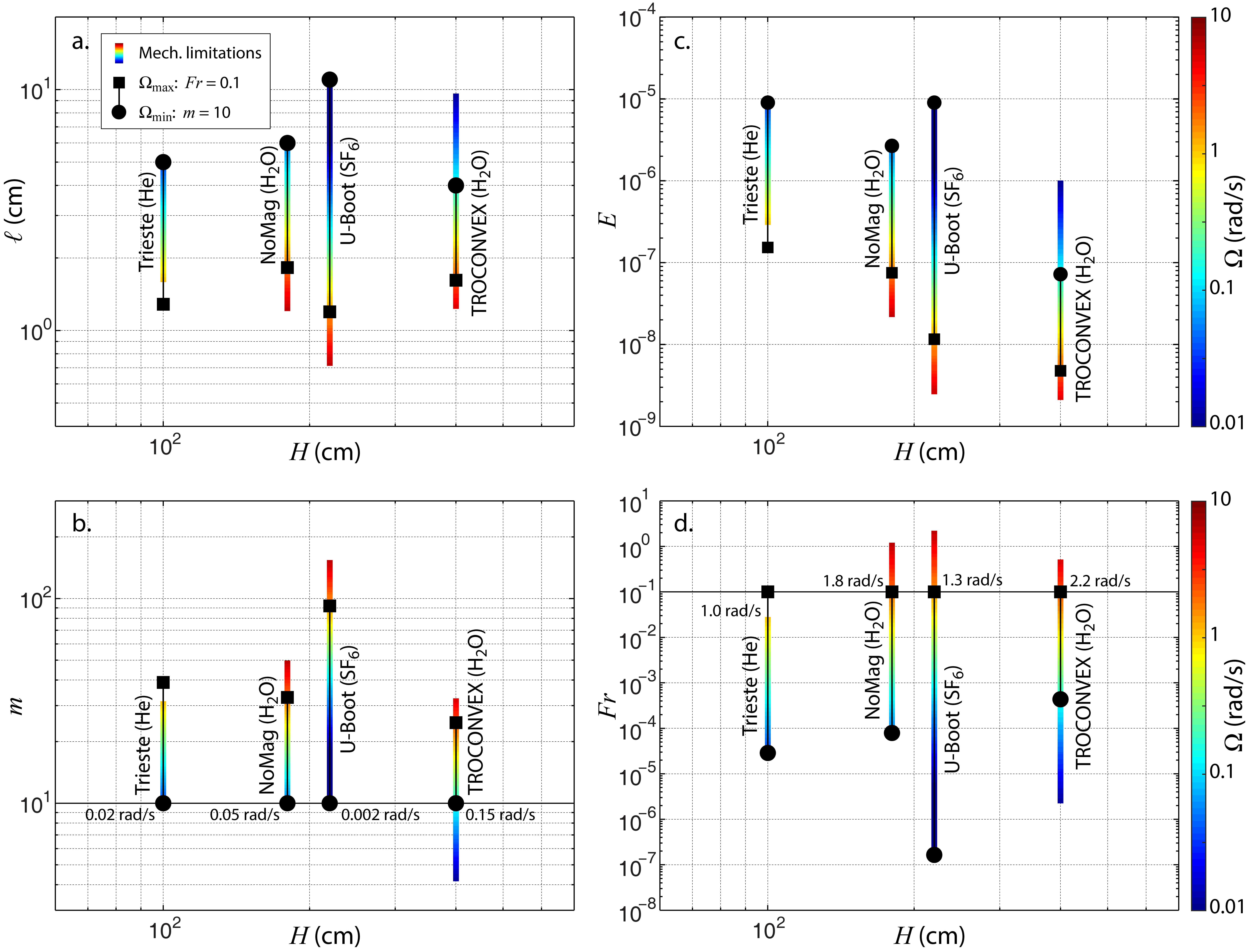}
\caption{\label{F:height} \footnotesize a) Flow structure width $\ell$, b) flow structure width ratio $m$, c) Ekman number $E$, and d) Froude number $Fr$ plotted versus tank height $H$ for four extreme rotation convection experiments. The tallest available tank in each experiment is used. The color bars specify the minimum and maximum achievable rotation rates based on mechanical limitations of each device. The filled black square represents the maximum rotation rate $\Omega_{\max}$ for which the Froude number $Fr = 0.1$, while the filled black circle represents the minimum rotation rate $\Omega_{\min}$ for which the flow structure width ratio $m = 10$. The values for $\Omega_{\min}$ in each experiment are given at the solid horizontal line in panel b, while the values for $\Omega_{\max}$ in each experiment are given at the solid horizontal line in panel d.
}
\end{figure*}

We define the flow structure width ratio $m$ as:
\begin{equation}
\label{eq:m}
m = D/\ell = c^{-1} E^{-1/3} \Gamma = c^{-1} \left(\frac{2 \Omega D^3}{\nu H} \right)^{1/3} \, ,
\end{equation}
where $\ell = 2.4 E^{1/3}$ is the typical onset width of convective rolls in $Pr \gtrsim  0.68$ fluids (see Appendix \ref{asym-regime}). We assume that for $m \geq 10$ sidewall effects do not dominate the bulk flows in a given experiment. This choice is somewhat arbitrary: while the thickness of the sidewall boundary layers scales as $E^{1/3}$ \eg{Greenspan63, Kunnen13}, the depth of sidewall effects on the bulk flow at low $E$ is not well-known.

Centrifugal effects contribute an upper bound on $\Omega$, and, thus, a lower bound on $E$. Centrifugation is parametrized via the Froude number \citep{Homsy69, Hart00, Curbelo14}:
\begin{equation}
\label{eq:Fr}
Fr = \frac{\text{centrifugation}}{\text{gravity}} = \frac{\Omega^2 D}{2 g} \, .
\end{equation}
In the case of high $Fr$, the centrifugal acceleration becomes significant relative to gravitational acceleration, causing denser parcels of fluid to travel radially outward. This leads to circulation patterns that are not found in the canonical rotating convection problem \eg{Hart00, Marques07}. To avoid the potential dynamical effects of centrifugation, we assign an upper limit of $Fr < 0.1$. Again, this choice is somewhat arbitrary: different studies have found different minimum $Fr$ values at which centrifugation first alters the flow \cf{Koschmieder67, Marques07}.

In  Figure \ref{F:height}, we compare the experimental device limitations to the limitations imposed by the $Fr < 0.1$, $m \geq 10$ constraints for each of the experimental setups shown in Figure \ref{F:imgs}. Of these devices, TROCONVEX can access the lowest Ekman number at $\approx 5 \times 10^{-9}$ due to its 4 m high tallest tank. However, the device also has the thinnest aspect ratio at $\Gamma = D/H = 1/10$. For a given experiment, the accessible $E$ range is:
\begin{equation}
\label{eq:Eratio}
\frac{E_{\max}}{E_{\min}} = \frac{\Omega_{\max}}{\Omega_{\min}} = \frac{\left( 8 g Fr_{\max} \right)^{1/2}}{c^3  \nu m_{\min}^3} \frac{D^{5/2}}{H} \, ,
\end{equation}
where $E_{\max}$, $E_{\min}$, $\Omega_{\max}$, and $\Omega_{\min}$ are defined in Table \ref{tab:constr}. As shown in Figure \ref{F:height}c, the large height and small diameter on the highest TROCONVEX tank cause its accessible $E$ range to be relatively small ($E_\mathit{\max}/E_\mathit{\min} = 15$), while the wide diameter of the U-Boot tank ($D=1.1$ m, $\Gamma = 1/2$) causes its accessible $E$ range to be relatively large ($E_\mathit{\max}/E_\mathit{\min} = 780$).

Apart from the physical capabilities of the experiment, the maximum heat transfer is also restricted by the dependence of the fluid properties on the temperature. A flow is considered to follow the Boussinesq approximation -- under which fluid properties do not change appreciably with temperature -- when the density difference that is driving the convection is small compared to the background fluid density \citep{Spiegel60, Busse67, Curbelo14, Horn14}:
\begin{equation}
\label{eq:Bouss}
\frac{\Delta \rho}{\rho_0} \ll 1 \rightarrow \gamma \Delta T \ll 1 \, ,
\end{equation}
where $\rho_0$ is the background density of the fluid and $\Delta \rho$ is the density perturbation. This enforces a separate upper bound on $\Delta T$. We have chosen the condition $\gamma \Delta T = 0.1$. However, some previous experimental studies have used more relaxed conditions such as $\gamma \Delta T = 0.2$ \eg{Niemela00}, while others have suggested additional criteria for ensuring Boussinesq conditions \citep{Busse67, Gray76}.

Figure \ref{F:RaE} shows the accessible $Ra$ versus $E$ ranges for the largest tanks on rotating convection experiments a) TROCONVEX, b) NoMag, c) Trieste device, and d) U-Boot. These ranges, indicated as green boxes, are bound in $E$ by the $m \geq 10$, $Fr < 0.1$ constraints. 
For the accessible $Ra$ ranges in the water experiments (panels a and b), the $\gamma \Delta T \leq 0.1$ condition is less restrictive than experimental limitations. Instead, in both cases, the maximum $\Delta T$ is determined by the maximum applicable heat flux while the minimum $\Delta T$ is determined by the precision of the temperature measurements. Gases tend to have greater thermal expansivities and are more likely to exceed Boussinesq limitations based on (\ref{eq:Bouss}). For example, $\Delta T < 10.6$ K is the maximum allowable temperature gradient for SF\textunderscript{6} in the U-Boot. These experiments are nevertheless capable of covering broad $Ra$ ranges by also varying the pressure.

Transition predictions are given as dashed lines in Figure \ref{F:RaE}. We were unable to find any formal predictions for the transition between plumes and geostrophic turbulence for $Pr > 3$ fluids, while the transition to nonrotating-style convection has several competing predictions in both $Pr >3$ and $Pr \lesssim 3$ fluids. One topic of interest for upcoming studies is to elucidate these transitions at lower $E$.

In $Pr > 3$ flows, the regime transitions between onset and nonrotating-style convection are far from evenly-divided: cells, columns, plumes, and geostrophic turbulence all take place relatively near onset. Both the TROCONVEX and NoMag experiments are also capable of characterizing the columnar regime at low $E$ and low $Ra$ values (high rotation rates and low $\Delta T$). They should comfortably access the plumes and geostrophic turbulence regimes, and associated transition scalings $Ra_{_\mathit{CP}}$ and $Ra_{_\mathit{PGT}}$, at low $E$ ranges.

The unbalanced boundary layers regime is predicted to cover a broad $Ra$ range for $Pr>3$ flows, with at least 2 decades separating $Ra_{_\mathit{GTU}}$ and $Ra_{_\mathit{UNR}}$. TROCONVEX and NoMag can both cover a decade of $Ra$ in this regime, and $Ra_{_\mathit{GTU}}$, though only NoMag will be able to access $Ra_{_\mathit{UNR}}$ and only at relatively high rotation rates and $E$ values.

\begin{turnpage}

\begin{figure*}
\centering
\includegraphics[width=0.9\linewidth]{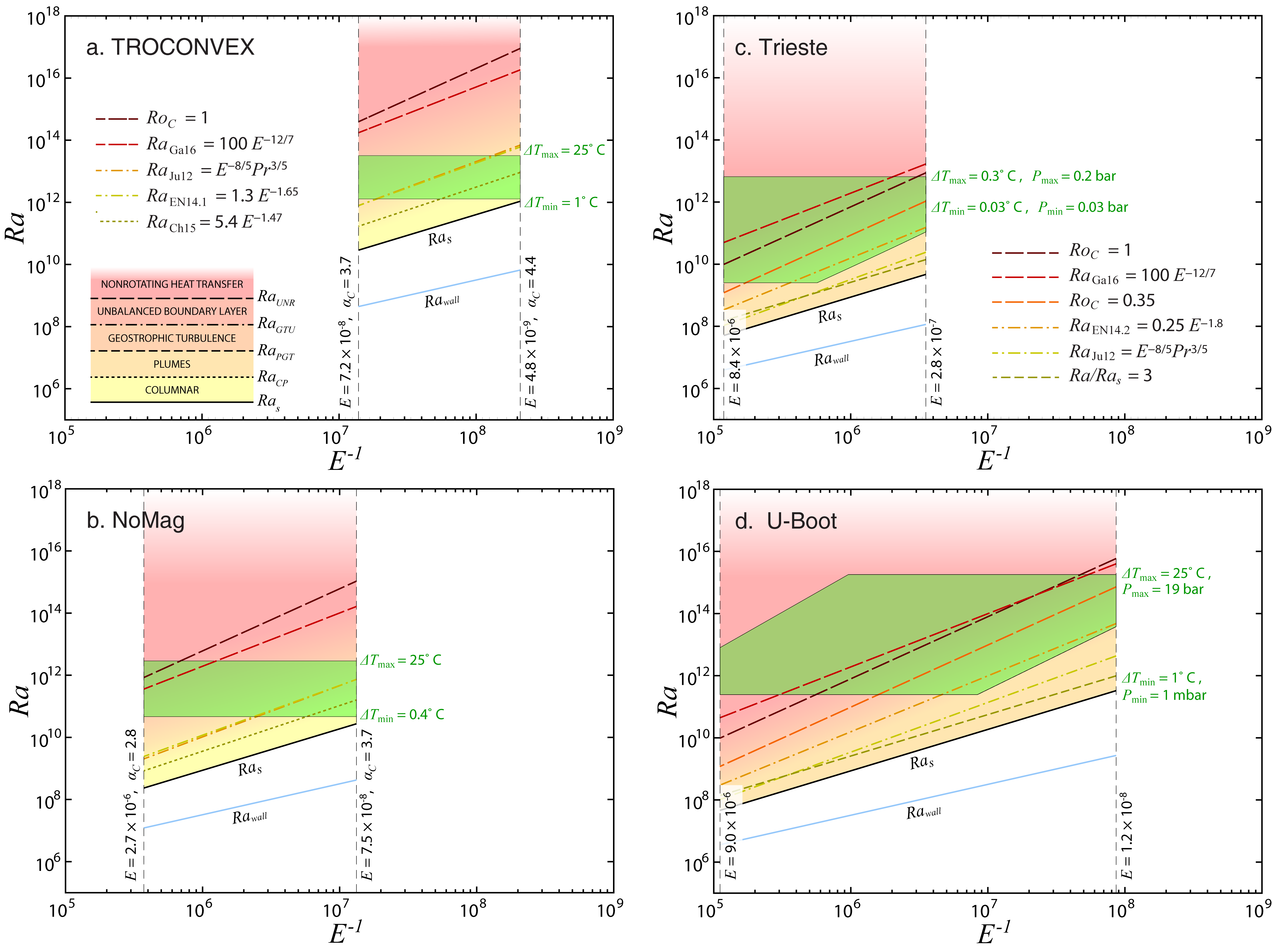}
\end{figure*}

\end{turnpage}
\clearpage

\captionof{figure}{
\linespread{1.0}\selectfont{}
\label{F:RaE} \footnotesize Rayleigh number $Ra$, plotted versus Ekman number, $E$, for the highest available tank size in each of the: a) TROCONVEX ($Pr=7$), b) NoMag ($Pr=7$), c) Trieste ($Pr=0.7$), and d) U-Boot ($Pr=0.7$) rotating convection experiments. The green box shows the range of $Ra$--$E$ space accessible to each experiment assuming fixed fluid properties. At the upper and lower $E$ bounds in panels a) and b), the slope of the $Nu$--$Ra$ scaling expected near the onset of convection is indicated, based on the $\alpha_{_C}$ trend from Figure \ref{F:slope}. Predicted regime transitions are plotted, with the line style indicating the type of transition and the line color indicating the specific prediction (following the legend; see Table \ref{tab:RaT}). Different background colors depict approximate locations of different flow regimes.}
 \noindent \hrulefill

For $Pr \lesssim 3$ fluids, $Ra_{_\mathit{GTU}}$ may occur very close to $Ra_{_\mathit{PGT}}$ or be separated by a decade in $Ra$, depending on which prediction is relevant. The U-Boot device should be able to test the \citet{Ecke14} prediction over at least a decade in $E$. Several competing predictions also exist for $Ra_{_\mathit{UNR}}$. Both the Trieste and U-Boot experiments should be able to thoroughly test this transition, with their ability to access both the unbalanced boundary layers and nonrotating-style regimes over nearly their entire accessible $E$ ranges and across several decades in $Ra$.

\section{Discussion}
\label{disc}

By expanding parameter coverage, upcoming studies will create opportunities to resolve the many unknown or conflicting transition predictions, scaling relations, and flow regime observations that exist in the rotating convection literature. We have compiled the key scaling predictions from multiple, independent studies to build a conceptual framework for studying the underlying physics in next-generation experiments.

Compiling results from numerous rotating convection studies reveals many gaps in our current understanding, such as a lack of predictions for the transition between plumes and geostrophic turbulence in $Pr > 3$ fluids and the existence of several competing predictions for the transition to nonrotating-style convection. 
Future experiments in unexplored parameter ranges will create opportunities to reconcile results from multiple, independent studies into a consistent physical model for extreme rotating convection.

Though we have outlined the regimes in terms of heat transfer, fully realizing such a model will undoubtedly also require length, time and velocity scale arguments. With the appropriate diagnostics, large-scale devices are poised to address important questions concerning these scales across multiple regimes. This knowledge should, in turn, lead to better design parameters for future experiments: the relationships between length and time scales determine the types of flow structures that can develop in a tank of given dimensions \citep{Nataf15}.

Based on existing predictions, each of the devices discussed is capable of reaching multiple behavioral regimes over broad parameter ranges. The water experiments are generally best suited toward accessing the regimes of geostrophic turbulence and plumes, and can access the columnar regime at the lower end of their accessible $E$ ranges. The gas experiments are best suited toward accessing the unbalanced boundary layers and nonrotating convection regimes over broad $E$ ranges. Open questions abound in every regime, and each of these devices should prove valuable for furthering our understanding of geostrophic convection.

While essential, understanding extreme geostrophic convection may only be a start toward understanding natural systems that are complicated by factors such as geometry, topographical effects, magnetic forces, compositional gradients, etc. Despite the complexity of these flows, though, there is evidence that purely hydrodynamic behaviors give critical insight into natural phenomena \eg{Kapyla11, Aurnou15}. As experiments foray into increasingly extreme conditions, they have a greater chance of encountering behaviors that are intimately linked to geophysics. The converse is also true: behaviors at moderate parameters which seem to underly natural phenomena can fail to scale up to more geophysical parameters \citep{Aurnou15, Cheng15}. In either case, laboratory studies are needed to form a concrete understanding of extreme flows in a real world setting. Developments in rotating convection are already bridging the gap between small-scale models and planetary-scale systems. Further advancing our understanding of rotating convection will require insights gained from both the current suite of experiments and from future experimental endeavors.

\begin{acknowledgments}
JSC and RPJK have received funding from the European Research Council (ERC) under the European Union's Horizon 2020 research and innovation programme (grant agreement n\textsuperscript{o} 678634). JMA and KJ thank the NSF Geophysics program for financial support. The authors thank Joseph Niemela and Robert Ecke for providing information about and images of the Trieste rotating convection experiment, and Ladislav Skrbek for providing the means to calculate the fluid properties of cryogenic helium. The authors also thank Stephan Weiss and Dennis van Gils for providing information about the U-Boot rotating convection experiment and a schematic of the device.
\end{acknowledgments}

\clearpage

\appendix

\section{Rotating convection regimes, scalings, and transitions}
\label{regimeapp}

\subsection{Regime predictions}
\label{asym-regime}

Between onset and $Ra/Ra_{s} \sim 2$, flow exists in the cellular regime \citep{Veronis59} (this regime is not marked separately on Figure \ref{F:NuRa}, as the heat transfer scaling does not change appreciably between this regime and the next \citep{JulienGAFD12}). For $Pr>3$, as $Ra/Ra_{s}$ increases, the `columnar' regime manifests \citep{Sprague06, Grooms10}. The bulk flow in this regime is dominated by quasi-steady convective Taylor columns, created by synchronization of the plumes emitting from the top and bottom boundary layers, and consisting of vortex cores surrounded by a shield of oppositely-signed vorticity \citep{JulienGAFD12}. In both the cellular and convective Taylor column regimes, the geostrophic balance between the Coriolis force and the pressure gradient is perturbed by viscous effects. This leads to narrow structures with a horizontal length scale of \eg{Zhang00, Stellmach04}:
\begin{equation}
\label{eq:ell}
\ell = c E^{1/3} H = c \left(\frac{\nu H}{2 \Omega} \right)^{1/3} \, ,
\end{equation}
where $c$ is a prefactor. While \citet{Chandra61} derives an asymptotic value of $c=4.8$ for the infinite plane layer, we use $c=2.4$ instead to account for the effects of Ekman pumping at $E>10^{-7}$ \citep{JulienTCFD98}. For $Pr \lesssim 3$ the steady columnar regime is not expected to manifest \eg{JulienGAFD12, KingAurnou13}.

At $Pr > 3$ and Rayleigh numbers in the vicinity of $Ra_{_\mathit{CP}}$ (marked by the short-dashed lines in Figure \ref{F:NuRa}), the shields surrounding the vortex cores in the columnar regime deteriorate and the flow enters the `plumes' regime. For $Pr \lesssim 3$, this regime develops directly out of cellular convection. The rising and falling plumes ejected from the boundaries are exposed to strong vortex-vortex interactions due to the deterioration of their shields, preventing them from synchronizing. This leads to structures which share the same horizontal length scale as columns and cells but which do not extend across the entire fluid layer \citep{JulienGAFD12}.

Around $Ra_{_\mathit{PGT}}$, shown as medium-dashed lines in Figure \ref{F:NuRa}, the `geostrophic turbulence' regime manifests. The plumes become confined close to the thermal boundary layers and the bulk of the fluid becomes dominated by strong mixing and vortex-vortex interactions \citep{JulienGAFD12}. 
Though small-scale turbulence is present, geostrophy still persists as the primary force balance and still imparts an effective vertical stiffness to the flow field.

Around $Ra_{_\mathit{GTU}}$, shown as dot-dashed lines in Figure \ref{F:NuRa}, geostrophy in the thermal boundary layers breaks down, leading to the theorized `unbalanced boundary layer' regime. The flow morphology here is not well-understood for $E \lesssim 10^{-7}$, and should be the subject of future studies.

Finally, around Rayleigh number $Ra_{_\mathit{UNR}}$, shown as long-dashed lines in Figure \ref{F:NuRa}, the `nonrotating-style heat transfer' regime is established as the flow field becomes effectively insensitive to Coriolis forces. For large enough $Ra$, the bulk of the fluid becomes nearly isothermal and the temperature gradients are almost entirely confined to the boundary layers.

\subsection{Heat transfer scaling ($\alpha$) predictions}
\label{alpha-regime}

\begin{figure}
\centering
\includegraphics[width=0.65\linewidth]{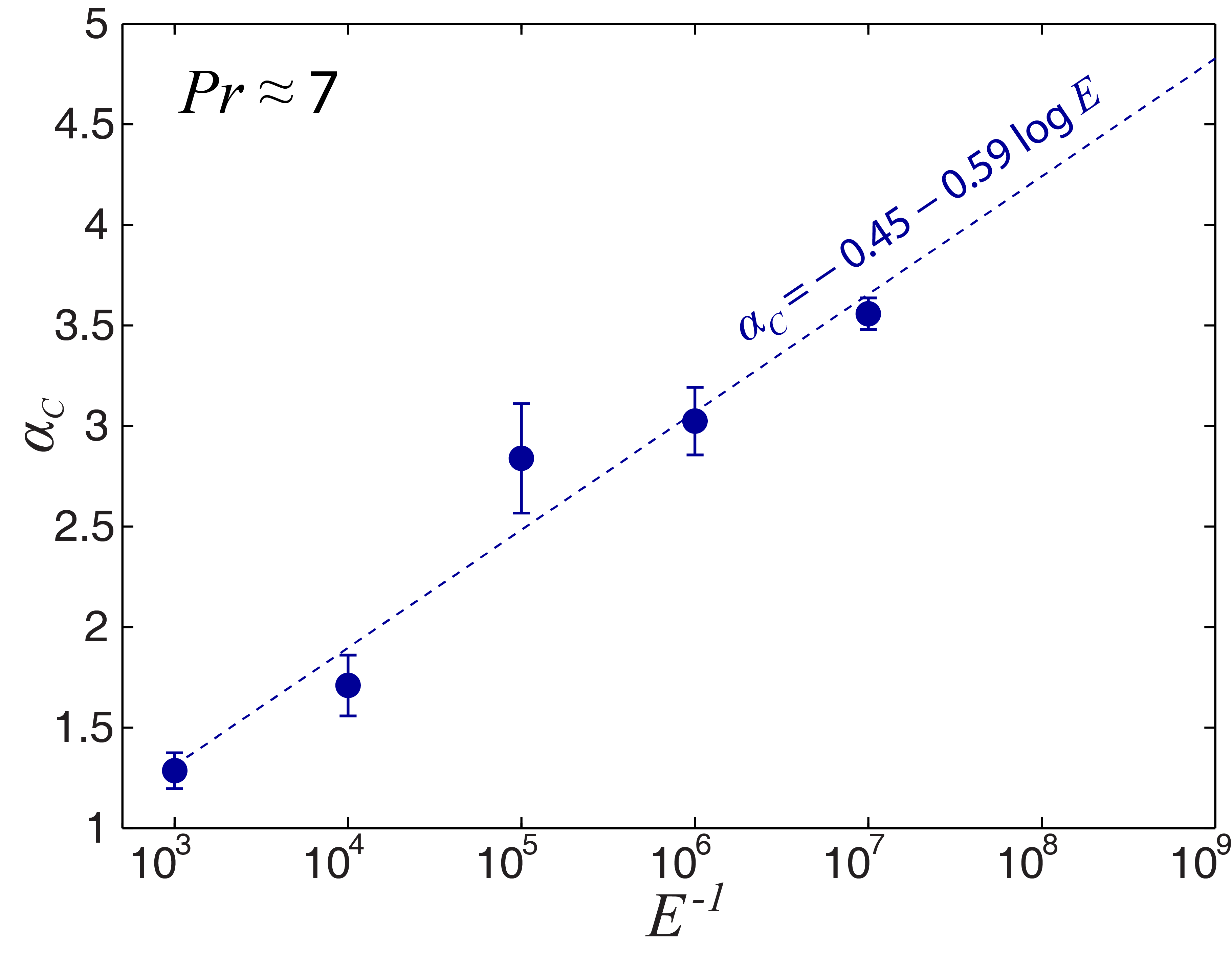}
\caption{\label{F:slope} \footnotesize Slope of Nusselt number versus Rayleigh number scaling in the columnar regime, $\alpha_{_C}$, plotted versus Ekman number, $E$, for $Pr \simeq 7$ laboratory-numerical rotating convection data from Cheng et al. (2015). The dashed line shows the best-fit slope between $\alpha_{_C}$ and $E$.}
\end{figure}

As $Ra/Ra_s$ increases from onset for fixed $E$, decreasing rotational control allows lateral mixing in the flow to increase. This causes $Nu$ to scale more weakly with $Ra$ in each subsequent regime at higher $Ra/Ra_s$. Here we will overview the $Nu$--$Ra$ scaling trends detected in previous rotating convection studies. 

In the columnar regime, the heat transfer follows steep $Nu \propto Ra^{\alpha_{_C}}$ trends, with $\alpha_{_C} \gtrsim 3$ for $E \lesssim 10^{-6}$ \citep{King12, Stellmach14, Cheng15}. The steepness of this trend is due to Ekman pumping effects, which greatly boost the heat transfer for a given thermal forcing \citep{Stellmach14, Kunnen16}. \citet{Julien16} theorize that Ekman pumping effects kick in above a threshold Rayleigh number:
\begin{equation}
\label{eq:Rathres}
Ra_\mathit{EP} \sim E^{-13/9} \, .
\end{equation}
This scaling implies that Ekman pumping should affect the flow immediately upon the onset of bulk convection for $E \gtrsim 10^{-9}$, although it is not known what the prefactor is for experiments. If we assume a prefactor of unity, then $Ra_\mathit{EP}$ lies below stationary onset $Ra_s$ for all of the experimental setups discussed.

Figure \ref{F:slope} demonstrates that the $\alpha_{_C}$ scaling exponent continues to steepen as $E$ decreases in $Pr \simeq 7$ laboratory and numerical rotating convection data from \citep{Cheng15}. The best-fit trend is given by:
\begin{equation}
\label{eq:alpha}
\alpha_{_C} = -0.45 - 0.59 \log E \, .
\end{equation}
In asymptotically-reduced simulations with parametrized Ekman pumping, \citet{Plumley16} find steep $\alpha_{_C}$ trends at $E=10^{-7}$, in agreement with laboratory and numerical results. As experiments become capable of reaching more extreme $E$ ranges, they will confirm whether the $Nu$--$Ra$ scaling law in the columnar regime continues to steepen as projected.

In the geostrophic turbulence regime, \citet{JulienPRL12} argue that the heat transport law should be independent of dissipation and predict that $Nu \sim Ra^{3/2}E^2 Pr^{-1/2}$ ($\alpha_{_{GT}} = 3/2$) \cf{Christensen06}. This is corroborated by \citet{Gastine16}, whose $Pr=1$ spherical shell rotating convection data follow a $Nu \sim Ra^{3/2}$ scaling for $Ra > 0.4 E^{-8/5}$.

In the nonrotating-style heat transfer regime, a variety of $Ra$ and $Pr$ dependent predictions exist for $\alpha_{_\mathit{NR}}$ \citep{Grossmann00, Ahlers09}. At high enough $Ra$, the temperature gradient becomes confined to the boundary layers. The heat transfer then becomes independent of the total height, leading to $\alpha_{_\mathit{NR}}=1/3$ \citep{Malkus54}. Water experiments have found evidence for this regime at $Ra \gtrsim 10^{10}$ \citep{Funf05, Cheng15}. At even higher $Ra$ values, \citet{Kraichnan62} and \citet{Spiegel71} predict that the boundary layers should become fully turbulent and the heat transfer should become independent of the diffusivities, leading to $\alpha_{_\mathit{NR}}=1/2$ (with logarithmic corrections). Some experiments find an increase in the heat transfer scaling at high $Ra$ \citep{Chavanne01, He12}, though this result is not universally supported \citep{Niemela00}. The majority of studies at $Ra < 10^{10}$ and $Pr \gtrsim 1$ instead find $\alpha_{_\mathit{NR}} \approx 2/7$ \citep{Rossby69, Chilla93, Glazier99}, theorized to be a nonasymptotic modification to the $\alpha_{_\mathit{NR}}=1/3$ scaling \eg{Castaing89, Shraiman90}. However, estimates of $Ra$ for planetary and stellar systems place them well beyond the expected parameter range for which the $\alpha_{_\mathit{NR}}=2/7$ scaling remains valid \eg{Schubert11}. 

In summary, a wealth of heat transfer scaling predictions exist for nonrotating and rotating convection. The relevance of each scaling to asymptotic settings, as well as their applicability to geophysical systems, remain open questions.

\subsection{Transition Rayleigh number ($Ra_T$) predictions}
\label{RaT-regime}

An empirical prediction for the columnar-to-plume transition is given by $Ra_{_\mathit{CP}} \sim 5.4 E^{-1.47}$, derived from laboratory and numerical $E=10^{-4}$ to $E=3 \times 10^{-8}$ rotating convection data in \citep{Cheng15}. It was determined by finding the intersection between the best-fit trend for the rotationally-controlled, steep $Nu$--$Ra$ scaling cases and the best-fit trend for nonrotating convection cases. Visualizations of the flow field in \citep{Cheng15} and thermal measurements in \citep{KingAurnou12} indicate that the breakdown of columnar structures into plume-like structures coincides with this intersection. For the $Ra$--$E$ ranges explored, the transition corresponds closely to the $Ra \sim 10 E^{-3/2}$ argument from \citep{King12}, where the $-3/2$ exponent describes the $Ra$--$E$ relationship for which the thickness of the Ekman boundary layer and the thickness of the thermal boundary layer become comparable.

Though no specific predictions exist for $Ra_{_\mathit{PGT}}$, in the asymptotically-reduced cases of \citep{JulienPRL12}, the heat transfer diverges from the geostrophic turbulence scaling $\alpha=3/2$ when $Ra \lesssim 3 Ra_{s}$. \citet{Ecke14} use this lower bound on the GT-scaling heat transfer as a broad estimate for $Ra_{_\mathit{PGT}}$ in $Pr > 3$ fluids.

\citet{JulienPRL12} predict that $Ra_{_\mathit{GTU}} \sim E^{-8/5} Pr^{3/5}$, where geostrophic balance in the thermal boundary layer breaks down in the asymptotic equations. \citet{Ecke14} find separate predictions for $Ra_{_\mathit{GTU}}$ depending on $Pr$: for $Pr = 0.7$ they argue that $Ra_{_\mathit{GTU}} \sim 1.3 E^{-1.65}$ while for $Pr=6$ they argue that $Ra_{_\mathit{GTU}} \sim 0.25 E^{-1.8}$. These empirical estimates assume that transitions take the form $Ra_T \sim E^{\chi}$ and are derived by determining the best-fit value of $\chi$.

\citet{Gilman77} predicts that the transition to nonrotating-style convection, $Ra_{_\mathit{UNR}}$, occurs when the system-scale buoyancy and Coriolis time scales become similar, or when $Ro_C \sim 1 \Rightarrow Ra \sim E^{-2} Pr$ \cf{Ecke14}. This prediction has been found to adequately describe the breakdown of large-scale zonal flows in spherical shell rotating convection simulations with free-slip boundary conditions \citep{Aurnou07, Gastine14} and the breakdown of the large-scale circulation in cylindrical rotating convection experiments \citep{Kunnen08}. \citet{Weiss11b} also find transitions in the heat transfer corresponding to constant $Ro_C$ values, but with an additional dependence on the aspect ratio $\Gamma$. \citet{Gastine16} empirically estimate $Ra_{_\mathit{UNR}} = 100 E^{-12/7}$ based on their spherical shell rotating convection simulations with non-slip boundary conditions. In the vicinity of this $Ra$ value, they find that all measurable quantities become indistinguishable from the nonrotating cases. Finally, \citet{Ecke14} find that the heat transfer becomes indistinguishable from nonrotating-style convection at $Ro_C \sim 0.35$ for $Pr \approx 0.7$.

These transition arguments have been compiled in Table \ref{tab:RaT}. Notably, no predictions have been made for $Ra_{_\mathit{GTU}}$ and $Ra_{_\mathit{UNR}}$ values. Larger datasets of more extreme rotating convection cases are needed, both to establish the validity of existing predictions and to develop new predictions for presently-unconstrained transitions.

\subsection{$\mathbf{Pr \ll 1}$ fluids}
\label{metals}

Liquid metal rotating convection follows a different set of predictions than those given for water and gas due to thermal diffusion operating on a far shorter time scale than viscous diffusion \eg{KingAurnou13}. Rotating convection onsets via oscillatory modes at \citep{Chandra61, JulienKnobloch98}:
\begin{equation}
\label{eq:Racrit_metal}
Ra_{o} \simeq 17.4 \left(\frac{E}{Pr} \right)^{-4/3} \, .
\end{equation}
The horizontal scale of these oscillatory structures are set by the thermal diffusivity \citep{Chandra61, JulienKnobloch98}:
\begin{equation}
\label{eq:ellmetal}
\ell_{o} \simeq c \left(\frac{E}{Pr} \right)^{1/3} H = c \left(\frac{\kappa H}{2 \Omega} \right)^{1/3} \, ,
\end{equation}
where $c=2.4$.

The flow structure width ratio is then given by:
\begin{equation}
\label{eq:mmetal}
m = D/\ell = c^{-1} \left( \frac{E}{Pr} \right)^{-1/3} \Gamma = c^{-1} \left(\frac{2 \Omega D^3}{\kappa H} \right)^{1/3} \, .
\end{equation}
Since the onset flow structures are significantly wider, we choose $m \geq 5$ as the minimum rotation rate condition for liquid metal experiments:
\begin{equation}
\label{eq:metomega}
\Omega_{\min} = \dfrac{62.5 c^3 \kappa H}{D^3} \, , \quad E_{\max} = \left(\dfrac{D}{5 c H} \right)^3 Pr \, .
\end{equation}
Table \ref{tab:RoMag} demonstrates the accessible ranges of $\ell$, $m$, $E$, and $Fr$ for the liquid gallium ($Pr \simeq 0.025$) experiment RoMag. The $Fr \leq 0.1$ condition allows for a minimum accessible $E$ of $2 \times 10^{-7}$, nearly an order of magnitude lower than existing liquid metal rotating convection studies in a cylinder.

\begin{table*}
\caption{\label{tab:RoMag} \footnotesize Rotation rate limits for the $H=0.5$ m tank on RoMag liquid gallium experiment ($Pr \simeq 0.025$), and corresponding flow structure widths $\ell$, flow structure width ratios $m$, Ekman numbers $E$, and Froude numbers $Fr$. Mechanical limitations are given by $\varOmega_{min}^{M}$ and $\varOmega_{max}^{M}$ while the $m \geq 5$ and $Fr \leq 0.1$ conditions are given by $\Omega_{\min}$ and $\Omega_{\max}$, respectively.}
\begin{ruledtabular}
\begin{tabular}{@{}lllll}
$\Omega$ (rad/s) & $\ell$ (cm) & $m$ & $E$ & $Fr$ \\
\hline
\noalign{\vskip 3pt}
$\varOmega_{min}^{M}$ = 0.04 & 10 & 2.0 & $1.5 \times 10^{-5}$ & $1.7 \times 10^{-5}$ \\
\noalign{\vskip 5pt}
$\Omega_{\min}$ = 0.70 & 4.0 & 5.0 & $9.3 \times 10^{-7}$ & $4.9 \times 10^{-3}$ \\
\noalign{\vskip 5pt}
$\Omega_{\max}$ = 3.1 & 2.4 & 8.3 & $2.0 \times 10^{-7}$ & 0.10 \\
\noalign{\vskip 5pt}
$\varOmega_{max}^{M}$ = 6.3 & 1.9 & 10.4 & $1.0 \times 10^{-7}$ & 0.40 \\
\end{tabular}
\end{ruledtabular}
\end{table*}

The onset of wall modes $Ra_{w}$, predicted by (\ref{eq:Raw}), occurs after oscillatory convection for $E \gtrsim 0.16 Pr^{4}$ ($E \gtrsim 6.3 \times 10^{-8}$ for $Pr \simeq 0.025$). Stationary onset $Ra_{s}$ occurs at still higher $Ra$ values, following (\ref{eq:Ras}). However, prior to stationary onset, \citet{Aurnou18} find that broad-band turbulence induced by wall and oscillatory modes has already manifested at $Ra_{b} \gtrsim 4 Ra_{o}$ in their liquid gallium experiments.

Liquid metals also behave differently from moderate $Pr$ fluids under nonrotating convection. At $Ra \gtrsim 2 \times 10^9$, \citet{Cioni97} find a $Nu \sim Ra^{2/7}$ scaling, consistent with $Pr \gtrsim 1$ results albeit with a different constant prefactor \citep{Scheel16}. At lower Rayleigh numbers ($\lesssim 5 \times 10^8$), though, studies find an $\alpha_{_\mathit{NR}} \simeq 1/4$ scaling where the heat transfer is controlled by inertially-driven, container-scale flows in the bulk rather than by viscous boundary layer processes \eg{Jones76, Cioni97, Horanyi99, KingAurnou13, Schumacher15}. \citet{KingAurnou13} find that their liquid gallium rotating convection cases conform to this nonrotating scaling for $Ra_{_\mathit{UNR}} \simeq \left( E^2/Pr \right)^{-1}$, corresponding to $Ro_C \simeq \mathcal{O}(1)$. 

\begin{figure}
\centering
\includegraphics[width=0.65\linewidth]{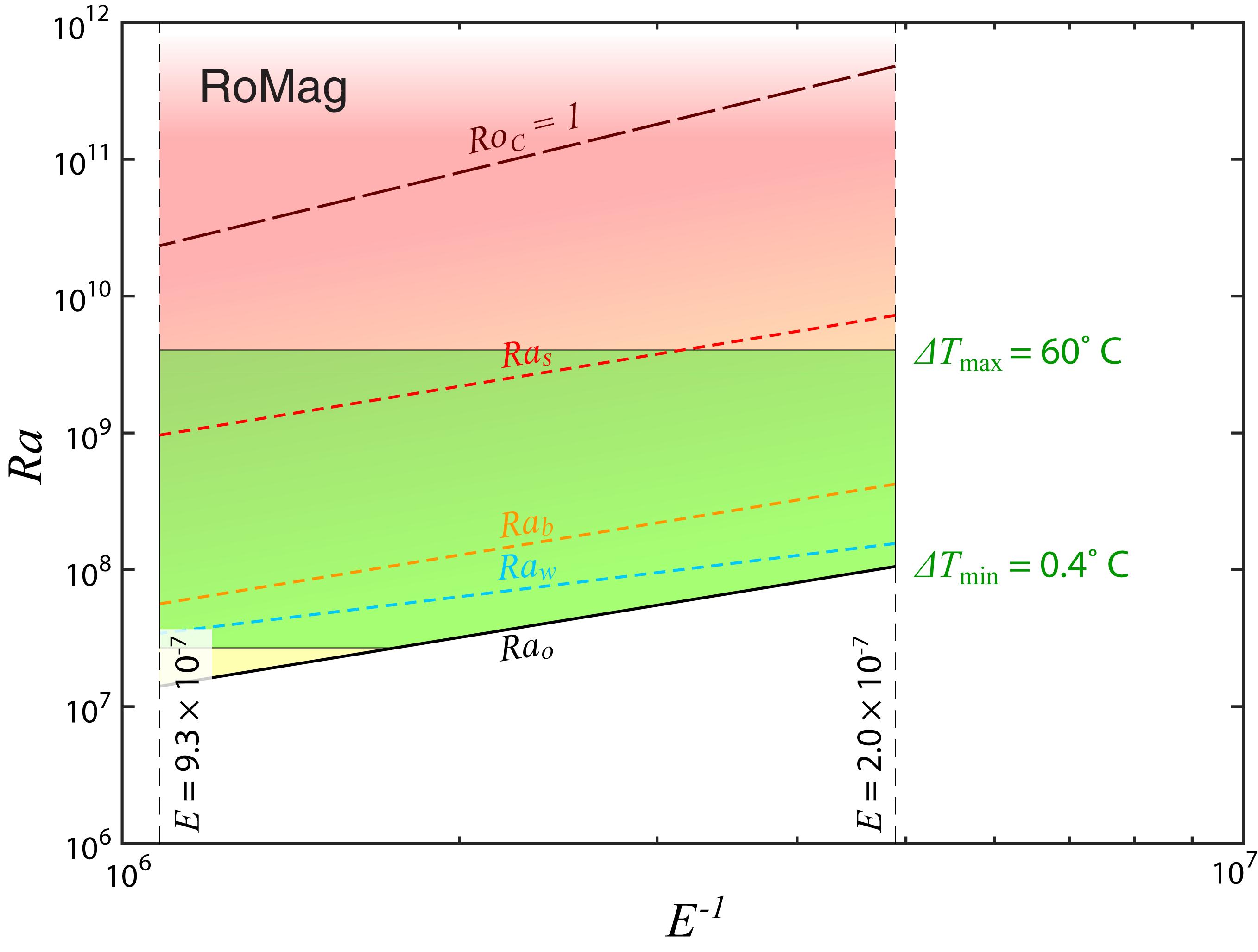}
\caption{\label{F:RaEGa} \footnotesize Rayleigh number $Ra$, plotted versus Ekman number, $E$, for the highest available tank size in the RoMag rotating convection experiment. The green box shows the range of accessible $Ra$--$E$ space, assuming fixed fluid properties. The onset occurs with oscillatory convection at $Ra_{o}$, with predicted regime transitions .}
\end{figure}

Figure \ref{F:RaEGa} shows the behaviors accessible to the 0.5 m high tank on RoMag based on the above predictions. In contrast to higher $Pr$ experiments, RoMag should be capable of delineating near-onset behaviors while encountering difficulty in accessing the nonrotating-style branch. RoMag should also be able to explore the transition to stationary convection for $E \gtrsim 3 \times 10^{-7}$.

While significant headway has been made in recent years toward understanding liquid metal rotating convection, open questions still abound with regard to the behaviors at $E \lesssim 10^{-6}$ and $Ra \gtrsim 10^{7}$. The $Pr \ll 1$ problem is generally relevant to a variety of planetary settings: in the outer cores of Mercury and the Earth and the metallic hydrogen layers of Jupiter and Saturn, $Pr$ values are estimated to be $\lesssim 10^{-1}$ \citep{Schubert11}. Further studies at even more extreme conditions in liquid metal may be essential for understanding such systems.

\clearpage

\section{Experiment information}
\label{exptinfo}
\begin{turnpage}

\begin{table}
\caption{\label{tab:app}\footnotesize Design properties and constraints for the experiments shown in Figure \ref{F:imgs} and discussed in Figures \ref{F:height} and \ref{F:RaE}. For entries with multiple values, the bolded quantity is the one used in our study. We include some references for devices used in previously published results. Minimum and maximum rotation rates $\Omega$ are determined by $m \geq 10$ and $Fr < 0.1$, respectively. Minimum and maximum temperature differences $\Delta T$ are determined by the measurement sensitivity and the Boussinesq limitation $\gamma \Delta T < 0.1$, respectively. An asterisk (*) in the $\Omega$ or $\Delta T$ category indicates that an experimental or diagnostic limitation is given instead because it is more restrictive than the corresponding theoretical constraint. The dagger ($^\dagger$) indicates that the minimum rotation rate uses an $m \geq 5$ condition instead of $m \geq 10$.} 
\begin{ruledtabular}
\begin{tabular}{@{}llllllll}
Experiment name & $H$ (m) & $D$ (m) & $\Omega$ (rad/s) & $\Delta T$ (\textdegree C) & input power (W) \\
\hline
\rule{-2.8pt}{2.8ex}
RoMag \cite{KingAurnou13} & 0.05, 0.1, 0.2, \textbf{0.5} & 0.2 & 0.70$^\dagger$ -- 3.1 & 0.4* -- 60* & 25 -- 5000 \\
Trieste \cite{Niemela00, Ecke14} & 1.0 & 0.5 & 0.02 -- 1.0* & 0.03* -- 0.2* & 0.001 -- 20 \\
NoMag & 0.1, 0.2, 0.4, 0.8, \textbf{1.8} & 0.2, \textbf{0.6} & 0.05 -- 1.8 & 0.4* -- 25* & 20 -- 1500 \\
U-Boot \cite{AhlersNJP09, Funf09} & 1.1, \textbf{2.2} & 1.1 & 0.002 -- 1.3 & 1* -- 10.6 & $\leq 3000$ \\
TROCONVEX & 0.8, 2.0, 3.0, \textbf{4.0} & 0.4 & 0.15 -- 2.2 & 1* -- 25* & 5 -- 2000 \\
\end{tabular}
\end{ruledtabular}
\\

\caption{\label{tab:app2}\footnotesize Fluid properties for the experiments shown in Figure \ref{F:imgs} and discussed in Figures \ref{F:height} and \ref{F:RaE}. In the `working fluid' column, the bolded item is the fluid for which properties are listed. $P$ gives the range of pressures used in each setup and $T$ the range of mean temperatures. The ranges for density $\rho$, coefficient of thermal expansion $\alpha_T$, kinematic viscosity $\nu$, thermal diffusivity $\kappa$, and Prandtl number $Pr$ are computed for the accessible ranges of $P$ and $T$ in the setups.}
\footnotesize
\begin{ruledtabular}
\begin{tabular}{@{}lllllllll}
Experiment name & working fluid & $P$ (bar) & $T$ (\textdegree C) & $\rho$ (kg/m\textsuperscript{3}) & $\alpha_T$ (1/K) & $\nu$ (m\textsuperscript{2}/s) & $\kappa$ (m\textsuperscript{2}/s) & $Pr$ \\
\hline
\rule{-2.8pt}{2.8ex}
RoMag & liquid Ga \cite{Aurnou18} & 1 & 35 -- 55 & 5900 & $1.3 \times 10^{-4}$ & $\left(3.4-3.7 \right) \times 10^{-7}$ & $1.3 \times 10^{-5}$ & 0.025 -- 0.028 \\
Trieste & cryogenic He \cite{CoolProp} & 0.03 -- 0.2 & -268.5 & 0.3 -- 2 & $3.7 \times 10^{-5}$ & $\left(0.54-3.7 \right) \times 10^{-6}$ & $\left(0.76-5.4 \right) \times 10^{-6}$ & 0.69 -- 0.72\\
NoMag & \textbf{water} \cite{Lide03}, air & 1 & 10 -- 50 & 990 -- 1000 & $\left(0.83-4.5 \right) \times 10^{-4}$ & $\left(0.5-1.3 \right) \times 10^{-6}$ & $\left(1.4-1.6 \right) \times 10^{-7}$ & 3.5 -- 9.4\\
U-Boot & \textbf{SF\textunderscript{6}} \cite{CoolProp}, He, N\textunderscript{2} & 0.001 -- 19 & 20 -- 35 & 0.0057 -- 160 & $10^{-2}$ & $\left(0.9-1.2 \right) \times 10^{-7}$ & $\left(1.2-1.3 \right) \times 10^{-7}$ & 0.75 -- 0.92\\
TROCONVEX & \textbf{water} \cite{Lide03}, air & 1 & 20 -- 80 & 970 -- 1000 & $\left(2-6 \right) \times 10^{-4}$ & $\left(0.34-1 \right) \times 10^{-6}$ & $\left(1.4-1.6 \right) \times 10^{-7}$ & 2.1 -- 6.9\\
\end{tabular}
\end{ruledtabular}
\end{table}

\end{turnpage}

\clearpage

\bibliographystyle{abbrvunsrtnat}
\bibliography{Bib_list}

\end{document}